\documentclass{aa}

\input{psfig.tex}
\def\gsim{\;\lower.6ex\hbox{$\sim$}\kern-7.75pt\raise.65ex\hbox{$>$}\;}
\def\lsim{\;\lower.6ex\hbox{$\sim$}\kern-7.75pt\raise.65ex\hbox{$<$}\;}
\begin{document}

%

%
\title{ Iron abundances from high-resolution spectroscopy of the open clusters
NGC 2506, NGC 6134, and IC 4651\thanks{
Based on observations collected at ESO telescopes under programme 65.N-0286,
and in part 165.L-0263} }

\author{
E. Carretta\inst{1,2},
A. Bragaglia\inst{2},
R.G. Gratton\inst{1},
\and
M. Tosi\inst{2}
}

\authorrunning{E. Carretta et al.}
\titlerunning{Iron abundances in open clusters}


\offprints{E. Carretta, carretta@pd.astro.it}

\institute{
INAF - Osservatorio Astronomico di Padova, Vicolo dell'Osservatorio 5, I-35122
 Padova, Italy
\and
INAF - Osservatorio Astronomico di Bologna, Via Ranzani 1, I-40127
 Bologna, Italy
  }

\date{Received ; accepted }

\abstract{This is the first of a series of papers devoted to derive the
metallicity of old open clusters in order to study the time evolution of the
chemical abundance gradient in the Galactic disk. 
We present detailed iron abundances from high resolution
($R\gsim 40,000$) spectra of several red clump and bright giant stars in the
open clusters IC 4651, NGC 2506 and NGC 6134. We observed 4 stars of NGC 2506,
3 stars of NGC 6134, and 5 stars of IC 4651 with the FEROS spectrograph at the 
ESO 1.5 m telescope; moreover, 3 other stars of NGC 6134 were observed with the UVES
spectrograph on Kueyen (VLT UT2). After excluding the cool giants near the red
giant branch tip (one in IC 4651 and one in NGC 2506), we found overall
[Fe/H] values of $-0.20\pm0.01$, $rms$ = 0.02 dex (2 stars) for NGC 2506,
$+0.15\pm0.03$, $rms$ = 0.07 dex (6 stars) for NGC 6134,  and $+0.11\pm0.01$,
$rms$ = 0.01 dex (4 stars) for IC 4651. The metal abundances derived from line
analysis for each star were extensively checked using spectrum synthesis of
about 30 to 40 Fe I lines and 6 Fe II lines. Our spectroscopic temperatures
provide reddening values in good agreement with literature data for these
clusters, strengthening the reliability of the adopted temperature and 
metallicity scale. Also, gravities from the Fe equilibrium of ionization agree
quite well with expectations based on cluster distance moduli and evolutionary
masses.
 }


\maketitle

\section{Introduction}
Our understanding of the chemical evolution of the Galaxy has tremendously
improved in the last decade, thanks to the efforts and the  achievements
both on the observational and on the theoretical sides. Good chemical
evolution models nowadays can satisfactorily reproduce the major
observed features in the Milky Way.
There are, however, several open questions which still need to be answered.

One of the important longstanding questions concerns the evolution of the
chemical abundance gradients in the Galactic disk.
The distribution of heavy elements with Galactocentric distance, as
derived from young objects like HII regions or B stars, shows a steep
negative gradient in the disk of the Galaxy and of other well studied spirals.

Does this slope change with time or not ? And, in case,
does it flatten or steepen ?

Galactic chemical evolution models do not provide a consistent
answer: even those that are able to reproduce equally well the largest set
of observational constraints predict different scenarios for early epochs.
By comparing with each other the most representative models of the time,
Tosi (1996) showed that the predictions on the gradient evolution range
from a slope initially positive which then becomes negative and steepens with
time, to a slope initially negative which remains roughly constant, to a
slope initially negative and steep which gradually flattens with time (see
also Tosi 2000 for updated discussion and references).

From the observational point of view, the situation is not much clearer.
Data on field stars are inconclusive, due to the large uncertainties
affecting the older, metal poorer ones. Planetary nebulae (PNe) are better
indicator, thanks to their brightness. PNe of type II, whose progenitors
are on average  2--3 Gyr old, provide information on the Galactic
gradient at that epoch and show gradients similar to those
derived from HII regions. However, the precise slope of the radial abundance
distribution, and therefore its possible variation with time is still subject
of
debate. In fact,  the PNe data of Pasquali \& Perinotto (1993),
Maciel \& Chiappini (1994) and Maciel \& K\"oppen (1994) showed gradient
slopes
slightly flatter than those derived from HII regions, but
Maciel, Costa \& Uchida (2003) have recently inferred, from a larger and
updated
PNe data set, a flattening of the oxygen gradient slope during the last 5--9
Gyr.

Open clusters (OC's) probably represent the best tool to understand whether and
how the gradient slope changes with time, since they have formed at all epochs
and their ages, distances and metallicities are more safely derivable than for
field stars. However, also the data on open clusters available
so far are inconclusive, as shown by Bragaglia et al. (2000) using the
compilation of ages, distances and metallicities listed by Friel (1995).
By dividing her clusters in four age bins, we find in fact no significant
variation in the gradient slope, but we do not know if this reflects
the actual situation or the inhomogeneities in the data treatment of
clusters taken from different literature sources. Inhomogeneity may lead, 
indeed, 
to large uncertainties not only on the derived values of the examined
quantities, but also on their ranking.
Past efforts to improve the homogeneity in the derivation of abundances,
distances or ages of a large sample of OC's include e.g., the  valuable works
by Twarog et al. (1997) and Carraro et al. (1998), but in both cases they had
to rely on literature data of uneven quality.  The next necessary step is to
collect data acquired and analyzed in a homogeneous way.

For this reason, we are undertaking a long term project of accurate
photometry and high resolution spectroscopy to derive homogeneously ages,
distances, reddening and element abundances in open clusters of various
ages and galactic locations and eventually infer from them the  gradient
evolution. Age, distance, and reddening  are obtained
deriving the colour-magnitude diagrams (CMDs) from deep, accurate photometry
and applying to them the synthetic CMD method by Tosi et al. (1991).
Accurate chemical abundances are obtained from high resolution spectroscopy,
applying to all clusters the same method of analysis and the same metallicity
scale.

Up to now we have acquired the photometry of 25 clusters and published the
results for 14 of them (see Bragaglia et al. 2004 and references therein,
Kalirai \& Tosi 2004). We have also obtained spectra for about 15 OC's, and have
already obtained preliminary results for a few of them.
Complete spectroscopic abundance analysis has been published so far only for
NGC 6819 (Bragaglia et al. 2001).

In this paper we present our results for three OC's, namely NGC 2506, 
NGC 6134, and IC 4651. Observational data are presented in Section 2,
while Section 3 is devoted to the derivation of atmospheric parameters, equivalent
widths and iron abundances. A check of the validity of our temperature scale
is derived from comparison with reddening estimates from photometry in Section
4. Spectral synthesis for all stars and its importance to confirm the validity
of our findings is discussed in Section 5. Finally, Sections 6 and 7 present 
a comparison with literature determinations, and a summary.

\begin{figure}
\psfig{figure=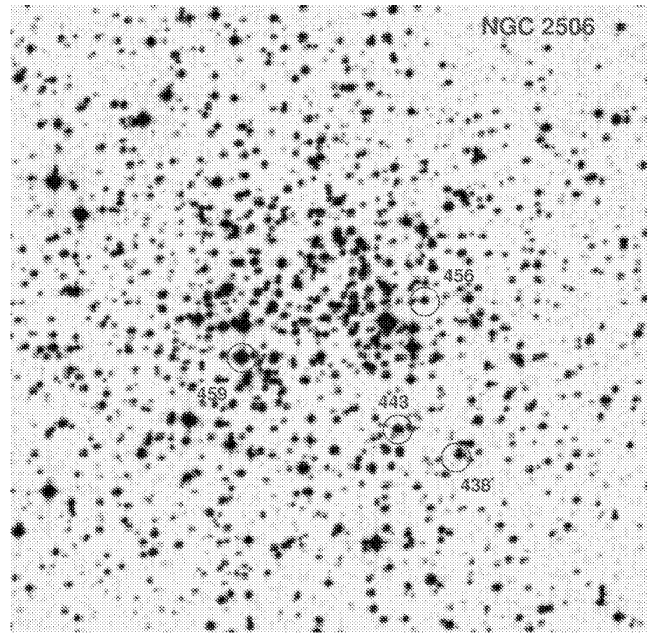,width=8.8cm,clip=}
\caption[]{Field of 10 $\times$ 10 arcmin$^2$ centered on NGC 2506, with the 4
stars observed with FEROS indicated by their numbers according to Marconi
et al. (1997)}
\label{f:field2506}
\end{figure}

\begin{figure}
\psfig{figure=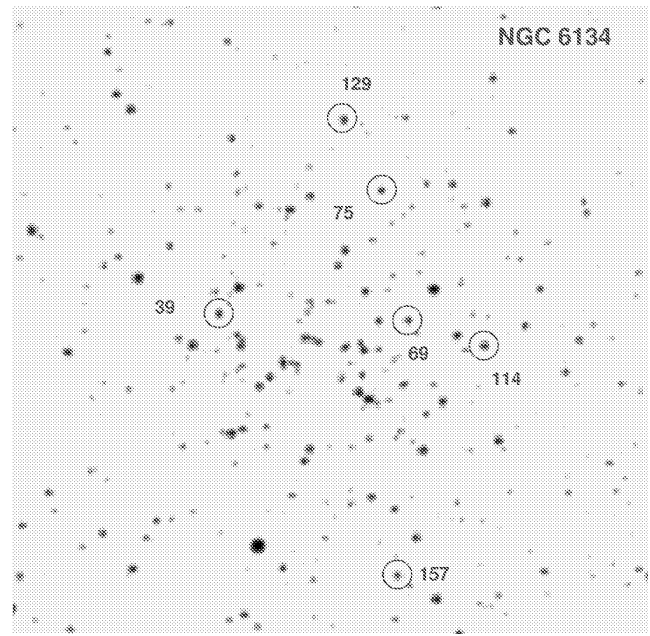,width=8.8cm,clip=}
\caption[]{Field of 10 $\times$ 10 arcmin$^2$ centered on NGC 6134, with the 6
stars observed with FEROS or UVES indicated by their numbers according to
Lindoff (1972).}
\label{f:field6134}
\end{figure}

\begin{figure}
\psfig{figure=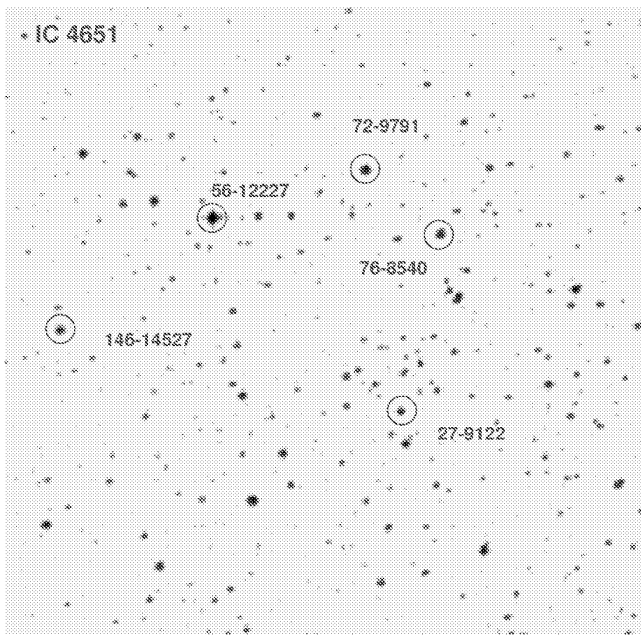,width=8.8cm,clip=}
\caption[]{Field of 10 $\times$ 10 arcmin$^2$ centered on IC 4651, with the 5
stars observed with FEROS indicated by their numbers according to Lindoff
(1972), and Meibom (2000).}
\label{f:field4651}
\end{figure}

\section{Observations}

Stars were selected on the basis of their evolutionary phases derived from the
photometric data. We mainly targeted Red Clump stars (i.e. stars  in the core
Helium burning phase) because, among the evolved population, they are the most
homogeneous group, and their temperatures are in general high enough that, even
at the high metallicity expected for open clusters, model atmospheres can well
reproduce the real atmospheres (this seems to be not strictly valid  for stars
near the Red Giant Branch - RGB - tip). We considered only objects for which
membership information was available  (except for one object in NGC 6134, which
however turned out to be member a posteriori), from astrometry (NGC 2506, Chiu
\& van Altena 1981), and/or radial velocity (NGC 2506: Friel \& Janes 1993,
Minniti 1995; NGC 6134: Clari\'a \& Mermilliod 1992; IC 4651: Mermilliod et al.
1995).

The three clusters were mainly observed with the spectrograph FEROS (Fiber-fed
Extended Range Optical Spectrograph) mounted at the 1.5m telescope in La Silla
(Chile) from April 28 to May 1 2000.
FEROS is bench mounted, and fed by two fibers (object + sky, entrance aperture
of 2.7 arcsec).  The resolving power is 48000 and the wavelength range is
$\lambda\lambda$ 370-860 nm. There is the possibility of reducing the data at
the telescope using the dedicated on-line data reduction package in the MIDAS
environment, so we could immediately obtain wavelength calibrated spectra.
Multiple exposures for the same star were summed.

Three additional stars in NGC 6134 were observed in July 2000 with UVES
(UV-Visual Echelle Spectrograph) on Unit 2 of the VLT ESO-Paranal telescope, as
a backup during the execution of another programme (169.D-0473).  These data
were acquired using the dichroic beamsplitter \#2, with the CD2 centered at 420
nm (spectral coverage $\lambda\lambda$ 356-484 nm) and the CD4 centered at 750
nm ($\lambda\lambda$ 555-946 nm). The slit length was 8 arcsec, and  the slit
width  1 arcsec (resolution of 43000 at the order centers). The UVES data were
reduced with the standard pipeline which produces extracted, wavelength
calibrated, and merged spectra.

\begin{figure*}
\hspace{1cm}
\psfig{figure=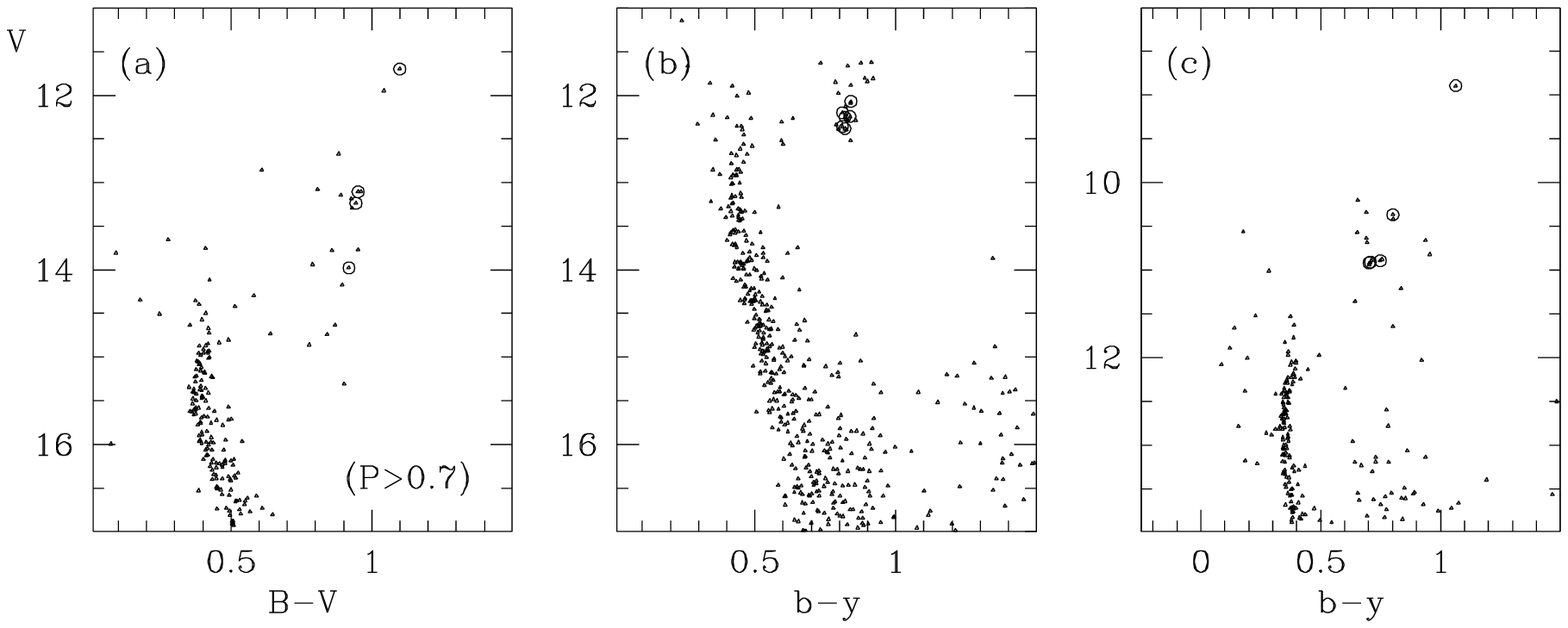,width=16cm,clip=}
\caption[]{CMD's for the three clusters, with the stars analyzed here indicated
by larger symbols. (a) NGC 2506: B and V taken from Marconi et al. (1997); only
stars with membership probability greater than 0.7 (Chiu \& van Altena 1981),
are shown. (b) NGC 6134: V, b-y taken from Bruntt et al. (1999). (c) IC 4651:
V, b-y taken from  Anthony-Twarog \& Twarog (2000).}
\label{f:cmd}
\end{figure*}

Finding charts for all observed targets are in Figure~\ref{f:field2506},
Figure~\ref{f:field6134} and Figure~\ref{f:field4651}. The evolutionary status
of observed stars is indicated by their position in the CMD, as shown in
Figure~\ref{f:cmd}.

Table \ref{t:reldata1} gives a log of the observations, and more information on
the selected stars is listed in Table~\ref{t:reldata2}, where values of the S/N
are measured at about 670 nm (for multiple exposures, these  values refer to
the final, co-added spectra).

\begin{table*}
\begin{center}
\caption{Log of the observations.
ID is taken from Marconi et al. (1997) for NGC 2506, from Bruntt et al. (1999)
for NGC 6134, and from Lindoff (1972) for IC 4651.
ID$_{BDA}$ is the identification number used in the BDA (Mermilliod 1995).
Coordinates are at J2000, and exposure time is in seconds.
F indicates FEROS spectra,
and U indicates UVES spectra. The P values given for the  stars
in NGC 2506 are membership probabilities.}
\begin{tabular}{rrccccrl}
\hline
ID  &ID$_{BDA}$  & Ra     & Dec          & Date obs.  & UT       &Exptime &Notes \\
\hline
\multicolumn{8}{c}{NGC 2506}\\
459 & 2122 & 8:00:05.84 & -10:47:13.33 & 2000-04-28 & 00:43:04 & 3600 & F, P=0.90\\
438 & 3359 & 7:59:51.79 & -10:48:46.51 & 2000-04-28 & 01:51:57 & 3600 & F, P=0.94\\
    &	   &	        & 	       & 2000-04-29 & 00:02:56 & 3600 & F     \\
443 & 3231 & 7:59:55.77 & -10:48:22.73 & 2000-04-29 & 01:06:30 & 3600 & F, P=0.91\\
    &	   &	        & 	       & 2000-04-29 & 02:12:58 & 3600 & F    \\
456 & 3271 & 7:59:54.06 & -10:46:19.50 & 2000-04-30 & 00:56:38 & 3600 & F, P=0.94\\
    &	   &	        & 	       & 2000-04-30 & 02:00:05 & 2400 & F    \\
    &	   &	        & 	       & 2000-05-01 & 01:34:51 & 2700 & F    \\
\multicolumn{8}{c}{NGC 6134}\\
404 & 114  &16:27:32.10 & -49:09:02.00 & 2000-04-30 & 08:04:09 & 2700 & F\\
    & 	   &		&              & 2000-04-30 & 08:55:42 & 3229 & F\\
929 &  39  &16:27:57.68 & -49:08:36.20 & 2000-05-01 & 07:45:47 & 1800 & F\\
    &	   &		&              & 2000-05-01 & 08:18:58 & 1800 & F\\
875 & 157  &16:27:40.05 & -49:12:42.20 & 2000-05-01 & 08:56:11 & 2400 & F\\
    &	   &		&              & 2000-05-01 & 09:39:14 & 2400 & F\\
428 &  75  &16:27:42.27 & -49:06:36.10 & 2002-07-18 & 01:25:11 & 2400 & U\\
421 & 129  &16:27:46.07 & -49:05:29.60 & 2002-07-19 & 23:36:34 & 1800 & U\\
527 &  69  &16:27:39.49 & -49:08:39.30 & 2002-07-20 & 00:09:57 & 1800 & U\\
\multicolumn{8}{c}{IC 4651}\\
  27&  9122& 17:24:50.13& -49:56:55.89 & 2000-04-28 & 08:05:27 & 1800  & F\\
  76&  8540& 17:24:46.78& -49:54:06.94 & 2000-04-28 & 08:39:43 & 1800  & F\\
  72&  9791& 17:24:54.22& -49:53:07.58 & 2000-04-28 & 09:15:05 & 1200  & F\\
  56& 12227& 17:25:09.02& -49:53:57.21 & 2000-04-28 & 09:39:59 &  600  & F\\
 146& 14527& 17:25:23.63& -49:55:47.11 & 2000-04-29 & 09:25:29 & 1500  & F\\
    &      &            &              & 2000-04-29 & 09:54:04 & 1500  & F\\

\hline
\end{tabular}
\label{t:reldata1}
\end{center}
\end{table*}

\begin{table*}
\begin{center}
\caption[]{Data for the observed stars in NGC 2506, NGC 6134, and IC 4651.
V, B-V are: CCD magnitudes from Marconi et al. (1997) for NGC 2506;
photoelectric measurements from Clari\'a \& Mermilliod (1992) for NGC 6134;
photoelectric
measurements from Eggen (1971), Lindoff (1972), and Jennens \& Helfer (1975), as
reported in the BDA, for IC 4651.
K is taken from 2MASS (Cutri et al. 2003). Stromgren y, b-y are: from Bruntt et al.
(1999) for NGC 6134, and Anthony-Twarog \& Twarog (2000) for IC 4651.
S/N has been measured at about 670 nm. Radial velocities (column 8, in km s$^{-1}$),
are heliocentric.}
\begin{tabular}{rcccccccl}
\hline
ID & V & B-V & K & y & b-y & S/N & RV & Phase \\
\hline
\multicolumn{9}{c}{NGC 2506}\\
459 &11.696 &1.100&  9.036&    -   &   -   & 110 &  +81.62 $\pm 0.6$ &RGBtip\\
438 &13.234 &0.944& 10.910&    -   &   -   &  85 &  +84.64 $\pm 0.3$ &clump \\
443 &13.105 &0.952& 10.791&    -   &   -   &  77 &  +84.66 $\pm 0.2$ &clump \\
456 &13.977 &0.919& 11.654&    -   &   -   &  35 &  +83.68 $\pm 0.3$ &RGB   \\
\multicolumn{9}{c}{NGC 6134}\\
404 &12.077 &1.310&  8.819&  12.072& 0.841 &  80 &$-$24.94 $\pm 0.6$ &clump \\
929 &12.202 &1.273&  9.042&  12.197& 0.811 &  56 &$-$25.34 $\pm 0.3$ &clump \\
875 &12.272 &1.268&  9.071&  12.251& 0.820 &  80 &$-$25.26 $\pm 0.2$ &clump \\
428 &12.394 &1.284&  9.208&  12.384& 0.820 & 200 &$-$26.34 $\pm 0.3$ &clump \\
421 &12.269 &1.314&  8.996&  12.245& 0.838 & 200 &$-$25.46 $\pm 0.3$ &clump \\
527 &  -    &  -  &  9.189&  12.359& 0.811 & 200 &$-$25.19 $\pm 0.3$ &clump \\
\multicolumn{9}{c}{IC 4651}\\
 27 &10.91  & 1.20&    -  &  10.896& 0.749 & 100 &$-$30.17 $\pm 0.6$ &clump \\
 76 &10.91  & 1.15&    -  &  10.915& 0.708 & 100 &$-$29.86 $\pm 0.3$ &clump \\
 72 &10.44  & 1.32&    -  &  10.371& 0.801 & 100 &$-$30.82 $\pm 0.2$ &RGB   \\
 56 & 8.97  & 1.62&  4.660&   8.899& 1.064 & 100 &$-$29.56 $\pm 0.3$ &RGBtip\\
146 &10.94  & 1.14&  8.340&  10.923& 0.702 & 100 &$-$27.88 $\pm 0.4$ &clump \\
\hline
\end{tabular}
\label{t:reldata2}
\end{center}
\end{table*}

\section{Atmospheric parameters and iron abundances}

\subsection{Atmospheric parameters}

We derived effective temperatures from our spectra, by minimizing the slope of
the abundances from neutral Fe I lines with respect to the excitation potential.
The gravities ($\log g$) were derived from the iron ionization equilibrium; to
this purpose, we adjusted the gravity value for each star in order to obtain an
abundance from singly ionized lines of iron 0.05 dex lower than the abundance from
neutral lines. This was done to take into account the same difference between Fe
II and F I in the reference analysis made using the solar model from the Kurucz
(1995) grid, with overshooting: the present adopted values are
$\log n$(Fe)=7.54 for neutral iron and 7.49 for singly ionized iron.

As usual, the overall model metallicity [A/H] was chosen as that of the model
atmosphere (extracted from the grid of ATLAS models with the overshooting
option switched on) that best reproduces the measured equivalent widths ($EW$'s).

Finally, we determined the microturbulent velocities assuming a relation
between $\log g$ and $v_t$. This was found to give more stable values than
simply adopting individual values of $v_t$ by eliminating trends in the
abundances of Fe I with expected line strengths. In fact, we found that the
star-to-star scatter in Fe abundances was reduced by adopting such a
relationship.

Our adopted atmospheric parameters for the three clusters are listed in
Table~\ref{t:atmpar}.

\begin{table*}
\begin{center}
\caption[]{Adopted atmospheric parameters and derived Iron abundances
for observed stars in NGC 2506, NGC 6134, and IC 4651; n indicated the number of
lines retained in the analysis.
In the last column we give for comparison the value of gravity derived from
photometric information.}
\begin{tabular}{rccccrccrccc}
\hline
Star   &  T$_{\rm eff}$ & $\log$ $g$ & [A/H] & $v_t$  & n &[Fe/H]I & $rms$ & n & [Fe/H]II & $rms$ &$\log ~g$\\
       &     (K)        &  (dex)     &  (dex)& (km s$^{-1}$) &&&&&&&(phot.)\\
\hline
\multicolumn{12}{c}{NGC 2506}      \\
459 & 4450 & 1.06 &$-$0.37  & 1.36  & 137& $-$0.37& 0.16 &10 &$-$0.37&0.17 & 1.74\\
438 & 5030 & 2.53 &$-$0.19  & 1.17  & 109& $-$0.18& 0.11 &13 &$-$0.18&0.15 & 2.68\\
443 & 4980 & 2.32 &$-$0.21  & 1.20  & 102& $-$0.21& 0.12 &13 &$-$0.21&0.09 & 2.60\\
456 & 4970 & 2.54 &$-$0.21  & 1.17  &  83& $-$0.21& 0.19 &10 &$-$0.21&0.14 & 2.95\\
\multicolumn{12}{c}{NGC 6134} \\
404 & 4940 & 2.74 & +0.11   & 1.14  & 128&  +0.11 & 0.16 &12 &  +0.11&0.12 & 2.60\\
929 & 4980 & 2.52 & +0.24   & 1.17  & 117&  +0.24 & 0.16 &11 &  +0.24&0.19 & 2.67\\
875 & 5050 & 2.92 & +0.16   & 1.12  & 126&  +0.16 & 0.14 &13 &  +0.16&0.15 & 2.73\\
428 & 5000 & 3.10 & +0.22   & 1.10  &  82&  +0.22 & 0.16 & 9 &  +0.23&0.14 & 2.76\\
421 & 4950 & 2.83 & +0.11   & 1.13  &  81&  +0.11 & 0.12 & 9 &  +0.11&0.13 & 2.68\\
527 & 5000 & 2.98 & +0.05   & 1.11  &  83&  +0.05 & 0.11 & 8 &  +0.05&0.12 & 2.71\\
\multicolumn{12}{c}{IC 4651}       \\
 27 & 4610 & 2.52 & +0.10   & 1.17  & 128&  +0.10 & 0.15 &10 &  +0.10&0.17 & 2.51\\
 76 & 4620 & 2.26 & +0.11   & 1.21  & 129&  +0.10 & 0.17 &11 &  +0.10&0.17 & 2.51\\
 72 & 4500 & 2.23 & +0.13   & 1.21  & 120&  +0.13 & 0.13 & 9 &  +0.13&0.21 & 2.25\\
 56 & 3950 & 0.29 & $-$0.34 & 1.46  &  89& $-$0.34& 0.17 & 6 &$-$0.35&0.15 & 1.23\\
146 & 4730 & 2.14 & +0.10   & 1.21  & 125&  +0.11 & 0.16 &11 &  +0.11&0.16 & 2.59\\
\hline
\end{tabular}
\label{t:atmpar}
\end{center}
\end{table*}

\subsection{Equivalent widths}

We measured the $EW$'s on the spectra employing an updated version of  the
spectrum analysis package developed in Padova and partially described in
Bragaglia et al. (2001) and Carretta et al. (2002). Measurements of Fe lines
were restricted to the spectral range 5500-7000~\AA\ to minimize problems of
line crowding and difficult continuum tracing blueward of this region, and of
contamination by telluric lines and possible fringing effects redward.

At these high metallicities and cool temperatures, continuum tracing is a major
source of uncertainty at the resolution of our spectra, hence we chose to use
an iterative procedure. For all clusters, the fraction of the 200 spectral
points centered on each line to be measured and used to derive the local
continuum level was first set to $1/4$. Following an extensive comparison with
spectrum synthesis of Fe I and Fe II lines (see below), we found that the
abundances for the stars in NGC 2506 and NGC 6134 (from FEROS spectra) were
much too large. This was explained by a too high continuum tracing  in
these spectra; for these two sets of data we finally adopted $1/2$ as the
fraction of spectral points to be used in the estimate of the local continuum.
With this new parameters, we performed again the automatic measurements of
$EW$'s. Then, the procedure strictly followed that described in Bragaglia et
al. (2001).

We used stars in the same evolutionary phase (in this case, clump stars) to
obtain an empirical estimate of internal errors in the $EW$'s. We performed the
cross-comparisons of the sets of $EW$'s measured for clump stars: 3 stars in IC
4651, 2 in NGC 2506 and 6 in NGC 6134 (in the latter case we  treated
separately the 3 stars observed with FEROS and the 3 from UVES). Assuming that
errors can be equally attributed to both stars in the couple under
consideration, we obtain typical errors in $EW$'s of 2.5 m\AA\ for clump stars
in IC 4651, 2.7 m\AA\ in NGC 2506, 3.6 m\AA\ and 3.1 m\AA\ in NGC 6134 (FEROS
and UVES spectra, respectively).

The classical formula by Cayrel (1989) predicts that, given the full width
half-maximum and the typical $S/N$ ratios (see Table~\ref{t:reldata2}) of our
spectra, the expected errors are 1.5, 1.8, 2.1 and 0.8 m\AA, respectively, in
the four cases.

The comparison with the observed errors shows that another source of error
(quadratic sum) has to be taken into account. This is likely to be 
uncertainties in the positioning of the continuum, an ingredient  neglected in
Cayrel's formula. If we consider for simplicity's sake lines of triangular
shape and the relationship between FWHM and central depth used by our
automatic procedure, we can estimate that the residual discrepancy can be well
explained by errors in the automatic continuum tracing at a level of 1\% for
clump stars in all clusters.

Sources of oscillator strengths and atomic parameters are the same of Gratton
et al. (2003) and discussion and references are given in that paper.

\subsubsection{Estimate of errors in atmospheric parameters}

\paragraph{Errors in effective temperatures:}
Uncertainties in the adopted spectroscopic temperatures can be evaluated from
the errors in the slope of the relationship between abundances of Fe I and
excitation potentials of the lines. Varying by a given amount the effective
temperature, these errors allow us to estimate a standard error of $88 \pm 9$,
$rms=29$ K (from 11 stars, excluding stars at the RGB tip).  Hence, we adopt a
standard error of about 90 K, corresponding to an average $rms$ of 0.018 dex/eV
in the slope.

This error ($\sigma^2_{\rm tot} = \sigma^2_{\rm rand.} + \sigma^2_{\rm syst.}$) 
for each individual star includes two contributions:
\begin{itemize}
\item a random term $\sigma^2_{\rm rand.}$, reflecting the various sources of
errors (Poisson statistics, read-out noise, dark current) of spectra, that
affects the measurements of the $EW$ 
\item a term $\sigma^2_{\rm syst.}$ due to a number of effects that
are systematic for each given line (e.g. the presence of blends, of possible 
features not well accounted for in the spectral regions used to determine the
continuum level, the line oscillator strength, etc.)
\end{itemize}

The first appears as a star-to-star scatter both in the slope
abundance/excitation potential and in the error associated to this slope.
Analogously, the second contribution shows up as a systematic uncertainty both
in the slope and in the associated error. In order to estimate the true random
internal error in the derived temperatures, this systematic  contribution has
to be estimated and subtracted.

We proceed in the following way. Let $j$ be the index associated to the stars
and N the number of stars; let $i$ be the index associated to the lines and
$M_j$ the total number of lines of neutral iron used in the $j-$th star.
We then have that the average variance of the distribution of the
abundances from individual lines for each star is:

$$ \sigma^2_{\rm tot} = \frac{\sum_j{\left[\frac{\sum_i{(x_{ij} -
\bar{x}_j)^2}}{M_j-1} \right]}}{N} $$

and as an estimate of the systematic contribution we may take: 

$$ \sigma^2_{\rm syst} = \frac{\sum_i{\left[\frac{\sum_j{(x_{ij} - 
\bar{x}_j)^2}}{Nj-1} \right]}}{M} $$

In our case, if we exclude from the computations the two tip stars and star 456
in NGC 2506 (the one with lower S/N), we obtain that  $\sigma_{\rm tot} =
0.141$ dex and $\sigma_{\rm syst.} = 0.110$ dex. Thus the  truly random
contribution to the standard deviation is (quadratic sum) 0.088 dex, hence  the
fraction of the standard deviation due to star-to-star errors is 0.088/0.141=
0.624. Hence, we may expect that the internal, random error in the temperatures
is actually $\sim 55$ K, to be compared with the observed value as derived from
independent methods (see below).

\paragraph{Errors in surface gravities:}
Since our derivation of atmospheric parameters is fully spectroscopic, a source
of internal error in the adopted gravity comes from the total uncertainty  in
T$_{\rm eff}$. 

To evaluate the sensitivity of the derived abundances to variations in the
adopted atmospheric parameters for Fe (reported in Table~\ref{t:sensitivity}),
we re-iterated the analysis of the clump star 146 of IC 4651 while varying each
time only one of the parameters of the amount corresponding to the typical
error, as estimated above. Considering the variation in the ionization
equilibrium given by a change of 90 K, $\partial$A/$\partial$T$_{\rm eff} =
0.143$ dex, and by a change of 0.2 dex, $\partial$A/$\partial g = 0.096$ dex, an
error of 90 K in temperatures translates into a contribution of 0.298 dex of
error in gravity.

From the above discussion, the contribution due to the random error in
temperature and to the error in gravities is then of 0.18 dex.

A second contribution is due to the errors in the measurements of individual
lines. If we assume as the error from a single line the average $rms$ of the
abundances from Fe I lines (0.14 dex, considering the same stars as above), 
again we must take into account only the random contribution, 0.088 dex, as
estimated above. Weighting this contribution by the average number of measured
line  (n=111 for Fe I and 12 for Fe II)  we obtain a random error of 0.03 dex
for the difference between the abundance from Fe II and Fe I lines. This
corresponds to another 0.06 dex of uncertainty in $\log g$, to be added in
quadrature to the previous 0.18 dex. The total random internal uncertainty in
the adopted gravity (quadratic sum) is then 0.20 dex that compares well with 
the estimate derived from the method described in the following.

\paragraph{Errors in microturbulent velocities:}
To estimate the proper error bars in the microturbulent velocity values, we
started from the original atmospheric parameters adopted for the clump star 146
in IC 4651. In this star the slope of the line strength $vs$ abundance
relation  has the same value of the quadratic mean of errors in the slope of
all stars, Hence it may be considered as a typical error. The same set of Fe
lines was used to repeat the analysis changing $v_t$\  until the 1$\sigma$
value from the slope of the abundance/line strength relation was reached. A
simple comparison allows us to give an estimate of 1$\sigma$ errors  associated
to $v_t$: they are about 0.1 km s$^{-1}$, thanks to the large number of used
lines of different strengths. Since we estimated that random errors are about
62\% of the error budget, the random internal error is then  0.06 km s$^{-1}$.

In order to estimate the total error, we have to recall that our final $v_t$
values are derived using a relationship between gravities and microturbulent
velocities from the former full spectroscopic analysis: $v_t = 1.5 -0.13 \times
\log g$. The $rms$ scatter around this relationship is 0.17 km s$^{-1}$, so
that the final error bar in microturbulent velocities derived through the
relation is 0.16 km s$^{-1}$, which includes a (small) component due to errors
in gravity and another component of physical scatter intrinsic to the different
stars.

Col. 7 of Table~\ref{t:sensitivity} allows to estimate the effect of errors in
the $EW$; this was obtained by weighting the average error from a single line
with the square root of the mean number of lines (listed in Col. 6 of the
Table) measured for each element. 

Notice that Cols. 2 to 5 in Table~\ref{t:sensitivity} are only meant to
evaluate the sensitivity of the abundances to changes in the adopted
atmospheric parameters. As discussed above, the actual random errors involved
in the present analysis are, more reasonably, 50 K in T$_{\rm eff}$, 0.2 dex in
$\log g$, 0.05 dex in [A/H] and 0.16 km s$^{-1}$ in the microturbulent velocity.
Moreover, the sensitivities in Cols. 2 to 5 of this Table are computed assuming
a zero covariance between the effects of errors in the atmospheric parameters.
In principle, this assumption is not strictly valid, since there are
correlations between different parameters: (i) effective temperature and
gravities are strictly correlated, so that for each 90 K change in T$_{\rm
eff}$ there is a corresponding change of $\sim 0.3$ dex in $\log g$, since
gravities are derived from the ionization equilibrium of Fe; (ii) there is a
correlation between $v_t$ and T$_{\rm eff}$, since lines of low excitation
potential tend to be systematically stronger than lines with high E.P. Thus, a
change in $v_t$ gives a change in effective temperatures as derived from the
excitation equilibrium. The sensitivity of this effect is not very large,
anyway, and it is about 20 K for each 0.1 km s$^{-1}$ change in $v_t$.

Taking into account these correlations in the computation of the total
uncertainty in the abundances, we have an error of 0.040 dex due to the
uncertainty of 0.16 km s$^{-1}$ in the $v_t$ value; moreover, another uncertainty of
0.032 dex comes from the uncertainty in temperature. If we neglect the other
contributions (that are small, anyway) from the quadratic sum we derive that the
abundance of an individual star bears a random error of 0.051 dex (listed in the
last Col. of Table~\ref{t:sensitivity}) that compares
very well with the observed scatter (0.053 dex) of individual stars around the 
mean value observed for each single cluster.

In the Table we omit in the last Column the total random error in the Fe II
abundance, since we force it to be identical to that of Fe I.

\begin{table*}
\begin{center}
\caption[]{Sensitivities of abundance ratios to errors in the atmospheric
parameters and in the equivalent widths}
\begin{tabular}{lccccrcc}
\hline
Ratio    & $\Delta T_{eff}$ & $\Delta$ $\log g$ & $\Delta$ [A/H] & $\Delta v_t$&$<N>$& $\Delta$ $EW$ & tot.\\
         & (+90 K)    & (+0.2 dex)      & (+0.1 dex)      & (+0.1 km s${-1}$) & &     & (dex)  \\
\\
\hline
& \multicolumn{7}{c}{Star IC4651-146 (Clump)} \\
\cline{2-8} \\
$[$Fe/H$]$I  &   +0.057& +0.011 &  +0.010 &$-$0.044&111&   +0.013& 0.051 \\
$[$Fe/H$]$II & $-$0.086& +0.107 &  +0.035 &$-$0.043& 12&   +0.040&       \\
\hline
\end{tabular}
\label{t:sensitivity}
\end{center}
\end{table*}

\subsubsection{Gravities from stellar models}

We can compare the gravities derived solely from the spectroscopic analysis
with what we obtain from the photometric information. We can compute gravities
using the relation:  $\log ~g = -10.607 + Log (M/M_\odot) -Log (L/L_\odot) +4
\times Log ({\rm T_{eff}}) $. For the temperatures, that enter the relation
both directly and indirectly through the bolometric correction (BC), we use our
derived ones, since they are  on the Alonso et al. (1996) scale, and are
therefore very close to what would be obtained from dereddened colours (see
next section). Values for masses are obtained reading the Turn-Off values on
the Girardi  et al. (2000) isochrones for solar metallicity at the generally
accepted ages for these clusters (1.7 Gyr and M=1.69 M$_\odot$ for NGC 2506 and
IC 4651,  0.7 Gyr and M=2.34 M$_\odot$ for NGC 6134). Adoption of any
reasonable different age or isochrone, and the fact that we are dealing with
(slightly) more massive stars, since they have already evolved from the main
sequence, would have a negligible impact on the final gravity. Absolute
magnitudes are computed using the literature distance moduli and reddenings:
(m-M)$_0$ = 12.6  (Marconi et al. 1997), 9.62 (Bruntt et al. 1999), and 10.15
(Anthony-Twarog \& Twarog 2000) for NGC 2506, NGC 6134, and IC 4651
respectively. The BC is derived from eqs. 17 and 18 in Alonso et al. (1999),
and we assume M$_{bol,\odot}$ = 4.75. Results of these computations are
presented in last column of Table  \ref{t:atmpar}. 

\begin{figure}
\psfig{figure=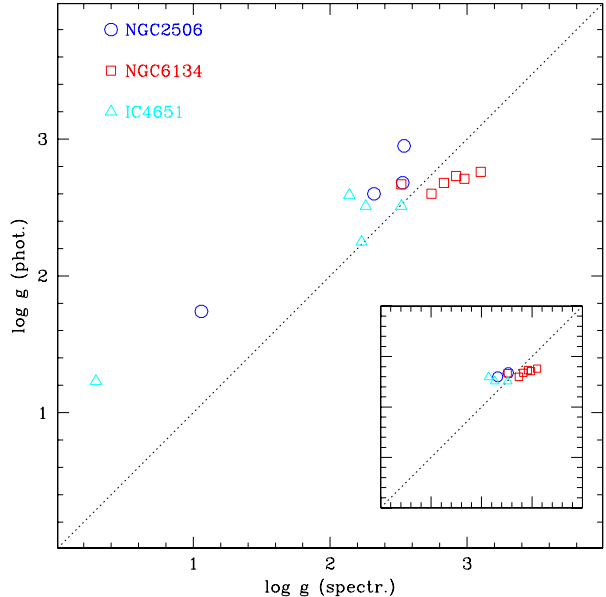,width=8.8cm,clip=}
\caption[]{Spectroscopic gravities as a function of the derived photometric 
gravities. Different symbols are used for the 3 clusters. In the inset, only the
clump stars and red giant stars are plotted, excluding the stars near the 
RGB-tip.}
\label{f:gravgrav}
\end{figure}

If we consider only the red clump stars, the average difference between the
gravities derived from photometry and spectroscopy for the whole sample is 
0.02 $\pm 0.07$ ($rms$ = 0.24, 12 stars). When we do the same for each cluster,
we find instead: +0.21 $\pm 0.07$ (rms = 0.09, 2 stars) for NGC 2506, $-$0.16
$\pm 0.07$ ($rms$ = 0.17, 6 stars) for NGC 6134, and  +0.18 $\pm 0.11$ ($rms$ =
0.22, 4 stars) for IC 4651. The internal error for each  cluster ($\sim$ 0.09
dex) is perfectly compatible with the internal errors in temperature and Fe
abundances. Instead, the cluster to cluster dispersion (0.21 dex) appears
larger.  
Many factors - and their combinations - could contribute to this
dispersion, among these: internal errors (0.09 dex), errors in distance moduli
(a conservative estimate of about 0.2 mag translates into about 0.08 dex) and
reddenings (almost negligible, since the values tied to the adopted distance
moduli are very similar to ours), in ages (hence masses, giving a small
contribution of less than about 0.03 dex even for a 20 \% age error), non
homogeneity of the data sources, small differences in the helium content of
these clusters not taken into account in the distance/age derivation
(responsible for less than about 0.05 dex in $\log g$),  systematic effects
somewhat tied to age (the two clusters with similar ages have also similar
differences in gravities, and opposite in sign to the younger one), etc. 
Moreover, plotting the derived values of photometric gravities $vs$ our
spectroscopic ones (Fig.~\ref{f:gravgrav}), we see a trend with  temperature:
clump stars of different clusters are segregated due to  differences in
metallicities and ages.   There is then a hint that the systematic effect
affecting the spectroscopic  gravities depends on the clump temperature. What
is the real physical effect  is not clear: it might be due to some Fe II lines
blended at lower temperature but clean at higher temperatures, or to the
atmospheric structure of stars   systematically varying with temperature.
Recently, various authors (see e.g.  Allende-Prieto et al. 2004, and references therein) showed  that similar problems are
also present in the analysis of dwarfs; however,  in our program stars the
effect is much more smaller and it does not  significantly affect our
conclusions on reddening, temperature and  metallicity scales. 

Given the small size of our sample, we cannot disentangle the above
possibilities, and we postpone a more in depth discussion until we have
examined more clusters in our sample. 

On the other hand, for stars at the RGB tip the agreement between the gravities
computed with the two methods is much worse, with gravities derived from
spectroscopy lower than those obtained from evolutionary masses.

In summary, when considering only clump stars, the agreement between
spectroscopic gravities (i.e. obtained through the ionization equilibrium) and
evolutionary gravities is on average good. This result leads to two relevant
conclusions: (i) if our gravity values are correct, also our temperatures must
be correct, hence our findings strongly support our temperature scale, which in
turn translates in a reliable $metallicity$ scale; (ii) the small differences
in gravity found with the two methods seem to hint convincingly against the
existence of large departures from the LTE assumption (see Gratton et al.
1999): high quality spectra coupled with reliable oscillator strengths from
laboratory experiments are able to provide solid Fe abundances for moderately
warm stars.

As far as the cooler RGB tip stars are concerned, the situation is clearly more
complex and less stable. In this case, we cannot exclude departures (even
relevant) from LTE, since the lower atmospheric densities do favour non-local
effects. Moreover, it is possible that the atmospheric structure of giant stars
near the RGB tip is not well reproduced by presently existing model atmospheres
of the Kurucz grid (see Dalle Ore 1993).  For these reasons we believe it is 
safer to concentrate on warmer objects for this kind of analysis, in order to
obtain more reliable results. In old OC's the red clump stars are the best
choice, given their temperatures and luminosities.

\section{Reddening estimates from spectroscopy}

Our effective temperatures are derived entirely from our spectra. As such, they
are reddening-free and this approach allows us to give an estimate of the
reddening toward the clusters which is independent of  the photometric
determinations. For this exercise, we used only the stars with T$_{\rm eff} \le$
4400 K, i.e. on the clump or near it.

First we collected all the available broad and
intermediate-band photometry for the analyzed stars. Johnson and
Str\"omgren photometry was taken from the references indicated in Table 2,
while JHK photometry
was obtained from the 2MASS survey (from the Point Source Catalogue of the 
All-Sky Data Release, found at {\tt http://www.ipac.caltech.edu/2mass/}) and
transformed to the TCS system.

We then adopted the colour-temperature transformations by Alonso et al. (1996)
and we entered our spectroscopic T$_{\rm eff}$ values in these relations
to obtain de-reddened colours. Comparison with the observed colours provides
an estimate of the $E(B-V)$ value.

Our final reddening values are derived as the weighted average of the reddening
values as given from individual colours (adopting $E(b-y)=0.72 E(B-V)$ and
$E(V-K)=2.75 E(B-V)$: Cardelli et al. 1989). As resulting errors, we adopted
the larger between the internal error and the spread in the values
obtained from individual colours.

\begin{table}
\begin{center}
\caption{Values for the reddening as determined in the present paper;
the three E(B-V) values for each cluster derived from individual colours,
i.e. from  (B-V), (b-y), and (V-K) respectively, are also given.
The last line gives the adopted reddening.}
\begin{tabular}{llll}
\hline
                    & NGC 2506         &      NGC 6134   &    IC 4651	   \\
\hline
${\rm [Fe/H]}$      & $-$0.20	       &       +0.15	 &   +0.11	   \\
T$_{\rm eff}$(ref.)   &  5000 K	       &      5000 K	 &  4500 K	    \\
E(B-V)$_{BV}$ &  0.058$\pm$0.016 & 0.355$\pm$0.005 & 0.087$\pm$0.022  \\
E(B-V)$_{by}$ & 		  & 0.340$\pm$0.006 & 0.082$\pm$0.015  \\
E(B-V)$_{VK}$ &  0.076$\pm$0.007 & 0.388$\pm$0.005 & 0.080     \\
\hline
E(B-V) adop.      &  0.073$\pm$0.009 & 0.363$\pm$0.014 & 0.083$\pm$0.011  \\
\hline
\end{tabular}
\label{t:reddenoi}
\end{center}
\end{table}

Our results for the three clusters are listed in Table~\ref{t:reddenoi}.
The agreement with previous literature data is quite good (see
Table~\ref{t:reddeliter}). Overall, the
differences are about 0.01 mag, with our values being slightly lower, on average
by 0.008 mag. This implies that our temperatures might be a little too low:
0.01 mag in $B-V$ translates into an offset of about 20 K in the
 derived
temperatures.

\begin{table}
\begin{center}
\caption{Literature values for E(B-V) and E(b-y).
SFD98=Schlegel et al. 1998; M+97=Marconi et al. 1997; K+01=Kim et al. 2001;
D+02=Dias et al. 2002 (catalogue); KF91=Kjeldsen \& Frandsen 1991;
CM92=Clari\'a \& Mermilliod 1992; B+99=Bruntt et al. 1999; AT+88=Anthony-Twarog
et al. 1988; ATT00=Anthony-Twarog \& Twarog 2000; M+02=Meibom et al. 2002}
\begin{tabular}{lll}
\hline
 E(B-V)          &   E(b-y)   & Reference \& notes\\
\hline
\multicolumn{3}{c}{NGC 2506} \\
  0.087          &                 & SFD98\\
 0-0.07          &              & M+97 ${\rm [Fe/H]} \simeq 0 ~{\rm to}~ -0.4$\\
  0.04$\pm$0.03  &                 & K+01 \\
  0.081          &                 & D+02 ${\rm [Fe/H]}$ = $-$0.37\\
\multicolumn{3}{c}{NGC 6134} \\
0.46$\pm$0.03    &                 & KF91 \\
0.35$\pm$0.02    &                 & CM92 ${\rm [Fe/H]}$ = $-$0.05\\
0.365$\pm$0.006  & 0.263$\pm$0.004 & B+99 ${\rm [Fe/H]}$ = +0.28 \\
0.395            &                 & D+02 ${\rm [Fe/H]}$ = +0.18\\
\multicolumn{3}{c}{IC 4651}  \\
0.086            &                 & AT+88 \\
0.241            &                 & SFD98 \\
0.086            & 0.062           & ATT00 ${\rm [Fe/H]}$ = +0.08\\
0.099            & 0.071           & ATT00 ${\rm [Fe/H]}$ = +0.12\\
                 & 0.071           & M+02 ${\rm [Fe/H]} \simeq +0.12$ \\
\hline
\end{tabular}
\label{t:reddeliter}
\end{center}
\end{table}

Our temperatures for individual stars have attached random errors of about
40-50 K; since we use from 3 to 6 stars in each cluster, we expect errors of
about 20-30 K per cluster, on average, hence 0.010-0.015 mag in the $E(B-V)$
value. Notice that the $\sim 90$ K estimated as the total internal error in
temperature (see above, Sect. 3.2.1) do include not only a random component due
to measurement errors, but also a systematic component due to the set of lines
used (e.g. some lines always giving too high or too low abundances).

On the other hand, such a good agreement with literature data strengthens the
reliability of our temperature scale, which is only 15 K (with 10-15 K of
uncertainty) too low with respect to that by Alonso et al. (1996) from IRFM.
This, in turn, might translate in our abundances being underestimated by
(only) 0.01 dex.

\begin{figure}
\psfig{figure=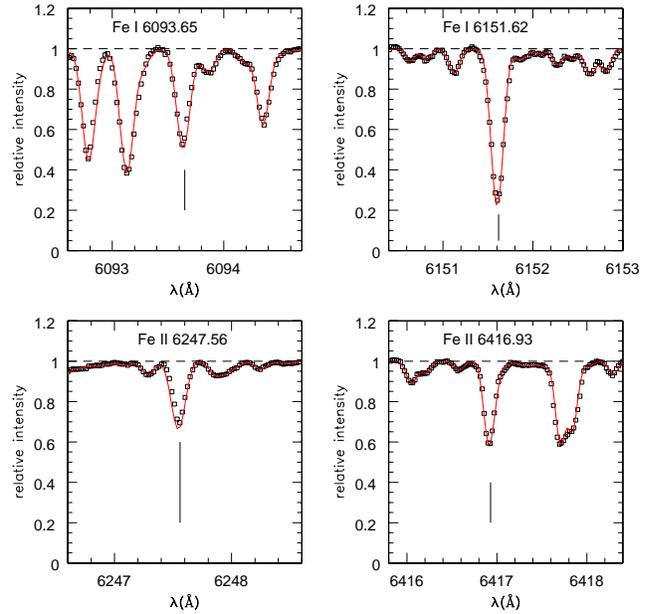,width=8.8cm,clip=}
\caption[]{Comparison between the observed spectrum of the field metal-rich 
giant star HR 3627 (open squares) and the synthetic spectrum computed by
optimizing the Kurucz line lists (solid lines). Small regions around 2 Fe I 
and 2 Fe II lines are shown.}
\label{f:fighr36b}
\end{figure}

\section{Analysis with synthetic spectra: a check}

\begin{figure}
\psfig{figure=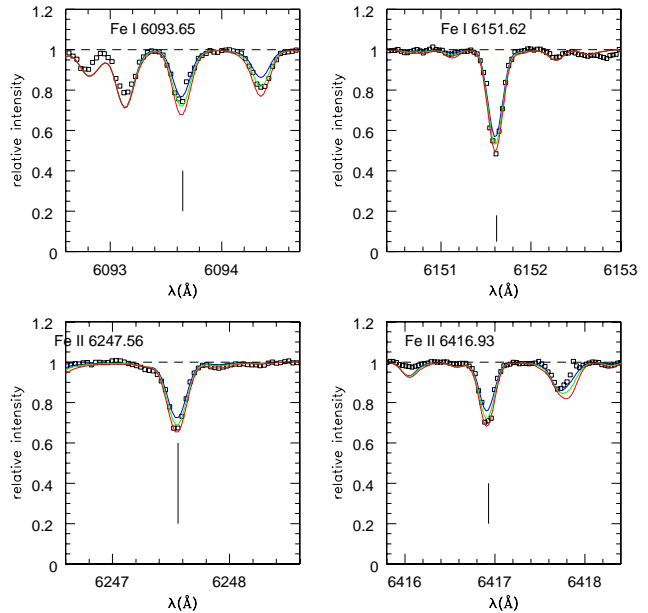,width=8.8cm,clip=}
\caption[]{Comparison between the observed UVES spectrum of star 421 in 
NGC 6134 (open squares) with synthetic spectra of two Fe I and two Fe II lines 
(solid lines).
The spectral synthesis was computed by using the atmospheric parameters
appropriate to the star and 3 Fe abundances: [Fe/H]=-0.2, 0.0 and +0.2 dex from
top to bottom, respectively.}
\label{f:fig61b}
\end{figure}

When dealing with high-metallicity, rather cool stars, the line crowding might
become rather difficult to treat simply with the standard method of line
analysis. In order to check our results, we devised a new procedure based on an
extensive comparison with spectrum synthesis:

\begin{itemize}
\item as a first step, line lists were prepared for  selected regions of 2~\AA\
centered on 44 Fe I lines and 9 Fe II lines chosen among those employed  in our
$EW$ analysis. These lines are in the wavelength range between 5500 and
7000~\AA, and were selected because they have very precise $\log gf$'s, and are
in comparatively less crowded spectral regions. \item For each region around Fe
lines we computed a  synthetic spectrum, starting from the Kurucz line list.
Our list was then optimized by comparing it with the very high resolution
($R\sim 140,000$), high signal-to-noise spectrum of star HR 3627, a metal rich
([Fe/H]=+0.35) field giant. This star was selected because its spectrum is very
line rich, due to the combination of low temperature and high metal abundance.
Hence we expect that a line list compiled from this spectrum would be adequate
to discuss even the most difficult cases among the stars in our sample. 
HR 3627 was observed with the SARG spectrograph mounted at the TNG, and
its atmospheric parameters were derived from a full spectroscopic analysis and
are: 4260/1.95/+0.35/1.20 (effective temperature, gravity, metal abundance and
microturbulent velocity, respectively). An example of the optimized synthetic
spectra for star HR 3627 is given in Figure~\ref{f:fighr36b}, where two Fe I
lines and two Fe II lines are showed, with the corresponding
synthetic  spectrum overimposed\footnote{Optimized line lists are available upon
request from the first author}.
\item After the line lists were optimized to well reproduce the spectrum of
star HR 3627 (in particular for contaminants like CN, with large contribution
in cool, metal-rich stars), we ran an automatic procedure that allows a
line-by-line comparison of the observed spectrum and the synthetic spectra of
the Fe lines.
\item For each program star, 7 synthetic spectra were computed for each Fe
line, varying the [Fe/H] ratios from $-$0.6 to +0.6 dex, in step of 0.2 dex.
\item The $EW$'s of these synthetic spectra were measured by using a local
pseudo-continuum that considers 2 small regions 0.8~\AA\ wide on each side of
the line. The measurements are then saved.
\item The same lines are measured exactly in the same way on the observed
spectra of the program stars.
\item A line-to-line comparison is made between the observed $EW$'s and those
measured on the synthetic spectra; the synthetic $EW$ that best matches the
observed one is then adopted as the one giving the right abundance;  examples
of the matches obtained for star 421 in NGC 6134 are shown in
Figure~\ref{f:fig61b} .
\item Finally, the resulting set of abundances are treated analogously to the
traditional line analysis, temperatures and microturbulent velocities are
derived by minimizing trends of abundances as a function of the excitation
potential and line strengths and so on. 
\end{itemize}

As mentioned above, the first iteration using this technique showed a
discrepancy with the values derived by line analysis for the NGC 2506 stars.
For stars in NGC 6134 the disagreement was found for stars with FEROS, but not
with UVES spectra, where the adopted fraction of high spectral points around
the lines used for the local continuum tracing was higher ($1/2$ rather than
$1/4$). This result leads us to conclude that the culprit was the location of
the continuum, traced too high in the FEROS spectra of NGC 6134 and NGC 2506.
We then repeated the $EW$ measurements by adopting $1/2$ as the fraction in the
routine. This time, the improvement was significant and the agreement is fairly
good, as shown in Table~\ref{t:sintesi}.

\begin{table}
\begin{center}
\caption[]{Comparison between Fe I abundances from spectrum synthesis and $EW$
analysis for observed stars in NGC 2506, NGC 6134, and IC 4651.}
\begin{tabular}{rrccrl}
\hline
Star & [Fe/H]I  & $rms$  & n  &   [Fe/H]I &Phase\\
     &   SS     &        &    &     $EW$  &\\
\hline
\multicolumn{6}{c}{NGC 2506}     \\	  
459  & $-$0.42  & 0.264  & 30 &   $-$0.37 &RGBtip\\
438  & $-$0.21  & 0.144  & 26 &   $-$0.18 &clump \\
443  & $-$0.15  & 0.079  & 19 &   $-$0.21 &clump \\
456  & $-$0.31  & 0.206  & 27 &   $-$0.21 &RGB   \\
\multicolumn{6}{c}{NGC 6134}     \\
404  &   +0.12  & 0.145  & 22 &     +0.11 &clump \\
929  &   +0.20  & 0.224  & 23 &     +0.24 &clump \\
875  &   +0.10  & 0.253  & 29 &     +0.16 &clump \\
428  &   +0.19  & 0.049  & 11 &     +0.22 &clump \\
421  &   +0.09  & 0.148  & 16 &     +0.11 &clump \\
527  &   +0.09  & 0.218  & 21 &     +0.05 &clump \\
\multicolumn{6}{c}{IC 4651}     \\
27   &   +0.16  & 0.200  & 25 &     +0.10 &clump \\
76   &   +0.07  & 0.141  & 19 &     +0.10 &clump \\
72   &   +0.18  & 0.172  & 21 &     +0.13 &RGB   \\
56   & $-$0.68  & 0.238  & 24 &   $-$0.34 &RGBtip\\
146  &   +0.18  & 0.198  & 23 &     +0.10 &clump\\
\hline
\end{tabular}
\label{t:sintesi}
\end{center}
\end{table}

The larger scatter in the Fe I abundances as obtained from spectrum synthesis
is likely due to the method used to measure the local pseudo continuum around
the Fe lines.

\section{Results and discussion}

Using the values derived for [Fe/H] only for the clump stars
(Table \ref{t:atmpar}) we have the following iron abundances:
[Fe/H] = 
$-0.20$ ($\sigma$= 0.02, 2 stars) for NGC 2506, 
 +0.15  ($\sigma$= 0.07, 6 stars) for NGC 6134, and 
 +0.11  ($\sigma$= 0.01, 3 stars) for IC 4651.

\begin{table*}
\begin{center}
\caption{Literature values for metallicity, and methods used. The two values
given for NGC 2506 by Marconi et al. correspond to the best reproductions of the
observed CMDs with synthetic ones based on the Padova tracks (Bressan et al.
1993) at Z=0.008 and 0.02; note that further unpublished analyses with
updated evolutionary tracks have shown that the solar solution should
be excluded. The two solutions for IC 4651 given by Anthony-Twarog
\& Twarog correspond to two different calibrations of the intrinsic b-y colour
versus metallicity.}
\begin{tabular}{llll}
\hline 
Reference & ${\rm [Fe/H]}$ & E(B-V) & method \\
\hline 
\multicolumn{4}{c}{NGC 2506} \\
this paper          & $-$0.20 & 0.073 & high-res sp.\\
Geisler et al. 1992 & $-$0.58 &      & Washington ph. \\
Friel \& Janes 1993 & $-$0.52 & 0.05 & low-res sp. \\
Marconi et al. 1997 & $\sim -$0.4 to 0.0 & 0.05 to 0 & synthetic CMD\\
Twarog et al. 1997  & $-$0.38 & 0.05 & low-res sp.+ DDO \\
Friel et al. 2002   & $-$0.44 & 0.05 & low-res sp. \\
\multicolumn{4}{c}{NGC 6134} \\
this paper          & +0.15 & 0.363 & high-res sp.\\
Clari\'a \& Mermilliod 1992 & $-$0.05 & 0.35 & Wash. + DDO \\
Twarog et al. 1997  & +0.18 & 0.35 -0.39 & DDO \\
Bruntt et al. 1999  & +0.28 & 0.365     & Str\"omgren \\
\multicolumn{4}{c}{IC 4651} \\
this paper          & +0.11 & 0.083 & high-res sp.\\
Twarog et al. 1997  & +0.09 & 0.11 - 0.12 & DDO \\
Anthony-Twarog \& Twarog 2000 & +0.077 & 0.086  & Str\"omgren \\
                              & +0.115 & 0.099  & Str\"omgren \\
Meibom et al. 2002  & $\sim$ +0.12   & 0.10  & CMD (Yale) \\
\hline 
\end{tabular}
\label{t:conf}
\end{center}
\end{table*}

\subsection{Comparison with other determinations}

None of these clusters has a previous metallicity measure based on high
resolution spectroscopy, but they have been the subject of many studies; we
present in Table \ref{t:conf} literature metal abundance, based on low
resolution spectroscopy, or photometric metallicity indicators (in DDO,
Washington and Str\"omgren filters), or comparison of observed CMDs with
theoretical isochrones/tracks. The reader is referred to the original papers
for detailed explanations, and we give here only a few notes on some of the
works.

Twarog et al. (1997) tried to derive in a homogeneous way the properties
of a large sample of OC's, and re-examined literature data to
find distances, reddenings and metallicities; for these  they
selected two methods, DDO photometry and the low resolution spectroscopy
of Friel \& Janes (1993), putting the two systems on the same scale.
Values cited in Table \ref{t:conf} come from their Tables 1 and 2.

Friel \& Janes (1993) collected low resolution spectra of giants
of a quite large sample of OC's; for an update, see Friel et al. (2002).

Marconi et al. (1997) used the synthetic CMD technique to determine at the same
time distance, reddening, age, and approximate metallicity of NGC 2506; they
employed three different sets of evolutionary tracks (Padova, Geneva, and
FRANEC) finding that, for Z = 0.008 and 0.01 ([Fe/H] $\simeq -0.4$ and $-$0.3),
they were able to reproduce the observed CMD, although the metal-poor solutions
were preferred. A similar method was employed by Meibom et al. (2002) for IC
4651, but they only considered the Yale isochrones.

Finally, Anthony-Twarog \& Twarog (2000) give two alternative solutions for IC
4651, based on different relations for the intrinsic Str\"omgren colours.

When we compare our findings with literature values, we find i) that our
abundance for NGC 2506 is generally much higher, ii) that NGC 6134 has
strongly discrepant determinations, and iii) that IC 4651 is in much better
agreement with past works.
We emphasize, though, that fine abundance analysis of high resolution, high
signal to noise spectra is the most precise method to measure the elemental
composition. Given also the tests done on our temperature and gravity scale,
we feel confident about the derived metallicities.

Finally, note that, adopting our metallicity  ([Fe/H] = +0.15), and the age
derived by Bruntt et al. (1999: 0.69  $\pm$ 0.10 Gyr), NGC 6134 appears almost
a twin of the Hyades for which Perryman et al. (1998) give [Fe/H]=+0.14 $\pm$
0.05, age 0.625 $\pm$ 0.05 Gyr, and distance modulus (m-M)$_0$ = 3.33 $\pm$
0.01. The similarity (although the clump stars in NGC 6134 are more numerous)
is confirmed by the absolute magnitude of the clump stars, which span a similar
range (M$_V$ $\sim$ 0.2 - 0.5) in the two OC's.

\section{Summary}

We derived precise metallicities for 3 open clusters (NGC 2506, NGC 6135 and IC
4651) from high resolution spectroscopy, using both the traditional line
analysis and extensive comparison with synthetic spectra. Our adopted
temperature scale from excitation equilibrium gives consistent values of
reddenings in very good agreement with previous, independent estimates of
$E(B-V)$ for the three clusters. This finding strongly supports the adopted
temperature scale and, in turn, the derived metallicity scale. The nice
agreement between spectroscopic and evolutionary gravities also indicates the
goodness of the adopted temperatures and leaves little space for the effect of 
possible departures from the LTE assumption. 
Our approach is then well suited to derive metal abundances of stellar
populations with a [Fe/H] ratio around solar.

\begin{acknowledgements}
{This research has made use of the SIMBAD data base, operated at CDS,
Strasbourg, France, and of the  BDA, maintained by J.-C. Mermilliod. This
publication makes use of data products from the Two Micron All Sky Survey,
which is a joint project of the University of Massachusetts and the Infrared
Processing and Analysis Center/California Institute of Technology, funded by
the National Aeronautics and Space Administration and the National Science
Foundation. This work was partially funded by Cofin 2000 "Osservabili stellari
di  interesse cosmologico" by Ministero Universit\`a e Ricerca Scientifica,
Italy. We thank the ESO staff at Paranal and La Silla (Chile) for their help
during observing runs, and P. Montegriffo for his precious software.}
\end{acknowledgements}

\end{document}